\documentclass[aps,prd,twocolumn,showpacs,groupedaddress]{revtex4-1}
 
\pdfoutput=1
\usepackage{amssymb}
\usepackage{amsmath}
\usepackage{hyperref}
\usepackage{graphics}

\begin{document}

\title{Stability and binding energy of small asymptotically Randall-Sundrum black holes}

 \author{Scott Fraser}
\email[Email address: ]{scfraser@calpoly.edu}
\affiliation{Department of Physics, California Polytechnic State University,  San Luis Obispo, California 93407, USA}

\author{Douglas M. Eardley}
\email[Email address: ]{doug@kitp.ucsb.edu}
\affiliation{Department of Physics, University of California, Santa Barbara, California 93106, USA}

\begin{abstract}
We study the binding of a small black hole to a positive-tension brane in the second Randall-Sundrum scenario (RS2) with orbifold symmetry.
We find that a small  black hole on the brane has substantial  binding energy to the brane, and is stable against escaping into the bulk.
This result can be applied in other models with an orbifold-symmetric brane.
We also find a novel static black hole, which is completely localized off the brane and is unstable against translations transverse to the brane.   
Our results are obtained analytically by applying a variational principle to black hole initial data.
This paper is the second in a series on asymptotically RS black holes. 
\end{abstract}

\pacs{04.50.Gh, 04.70.Bw}

\maketitle

\section{Introduction}

There is ongoing interest in  
   the 
Randall-Sundrum (RS)     models  \cite{RS1,RS2},
where  our observed universe is a brane surrounded by a 
higher-dimensional 
 anti-de Sitter  (AdS)  bulk.
 If the higher-dimensional Planck
energy is of order TeV,
 an exciting prediction in the RS1 model \cite{RS1} is the possible  production of
small  black holes at TeV scale collider energies     \cite{Banks-Fischler, *Giddings},
and
  LHC experiments  
  \cite{LHC-RS-extra-dimensions, LHC-RS-black-holes} continue to test this hypothesis. 
  
 In such models,    standard model particles are   confined to the brane,
   while  gravity (described by spacetime curvature)    propagates  in the bulk.  
A black hole is a purely gravitational object,  so a
  natural question
is whether a  black hole    on a brane could
 escape into the bulk.
 In the RS models, a brane  has    orbifold  
 symmetry: mirror  points across the brane are identified.  
Without orbifold symmetry,
a small black hole  can    escape 
 \cite{recoil, *domain-walls}, possibly  
    pinching off some of the brane
\cite{flachi-tanaka-BH-escape,*flachi-BH-escape-tension}.
Orbifold  symmetry would appear to
forbid  such pinching off, but the question of escape has generally remained open;
in this paper, we   will   give different arguments against escape   than       in \cite{ADD-vs-RS}.
For a large black hole on the brane,
 AdS/CFT arguments     
 suggest  that   the black hole    may
classically evaporate by
  emitting
gravitational waves \cite{emparan-fabbri-kaloper-BH-holograms}
or     smaller black holes
  \cite{tanaka-BH-evaporation, *fitzpatrick-randall-wiseman}.

The RS1 model \cite{RS1}  has two branes of opposite tension, with
our universe  on the negative-tension brane.    
 In the RS2 model \cite{RS2}, 
our  universe resides on  the positive-tension brane,
with the negative-tension brane  removed to infinite distance.
Perturbations of RS2   reproduce
 Newtonian gravity  at large distance  on the brane, while in
 RS1 this requires a mechanism to stabilize the interbrane distance \cite{Tanaka}.
In  RS2,   solutions for static black holes    on the brane   have been found  numerically,   for both
    small black holes
\cite{Kudoh-smallBH-1, *Kudoh-smallBH-2, Kudoh-smallBH-6D} and large  
 black holes  \cite{Figueras-Wiseman, *Abdolrahimi}, compared to the AdS curvature length.
The only known   analytic black hole
solutions
are the static and stationary  solutions   \cite{ehm, ehm-2} in a lower-dimensional version of RS2.
 An exact solution for a large black hole on the brane was also found in \cite{nonempty-bulk} in a generalized RS2 setup with matter in the bulk.

In this paper, 
we   examine
the  binding of small  black holes to a positive-tension brane with orbifold symmetry in  RS2. 
No exact solutions are known for these black holes, and numerical methods have been essential to   study them. 
Here, we  take a different  approach   using  our   variational principle  \cite{paper-1-first-law}
for  black holes in RS2:
\begin{equation}
\parbox[b]{3in}{
{\it Initially   static   initial data that 
extremizes the mass     is  initial data for
a static    black hole, for   variations
at fixed  apparent horizon area $A$, 
  AdS curvature length $\ell$, cosmological constant $\Lambda$, brane tension   $\lambda$
and  asymptotic   warp factor     $\psi_0$ on the brane.}}
  \label{vp}
\end{equation}
Our approach is    analytical.  There has also been    some numerical work    \cite{floating-BH-initial-data} using extrema  
   in a detuned RS2 setup.

This paper is organized as follows.
We derive a   general binding energy result 
in section \ref{binding energy},
and review the   RS models
in  section \ref{RS-review}.
 We formulate our
  initial data  in section
\ref{chap:setup-initial-value}, and    solve
 the resulting boundary value problem 
 in  the following  two different regimes.
 In section \ref{ch:Misner}, for 
  small  black holes
 on or near  the brane,
our variational principle reproduces the well known
 static   braneworld black hole, which we show
 is translationally stable and has  large   binding energy to the brane.
In section \ref{ch:R-S},
for
  small  black holes
farther from the brane, our variational principle locates
a new static black hole, which is 
translationally unstable.   
In section  \ref{bounds}, we examine the  energy and length scales for which our results are valid, and the related  phenomenology.
 We conclude in section \ref{conclusions}.

We
work   primarily on the
orbifold  (one side of the brane)
  in $D=5$ spacetime dimensions.
Spatial geometry
has metric $h_{ab}$  and covariant derivative $D_a$.
A spatial boundary  has  
  metric  $\sigma_{ab}$, extrinsic curvature $k_{ab}=h_a{}^c D_c n_b$,
   and outward unit normal     $n_a$.
    We   often use $\simeq$ for approximations at leading order, and  often refer to
the black hole rest mass $M_A$,  defined  by   horizon  area $A$   and
     area  $\omega_{D-2}$ of the $(D-2)$-dimensional unit sphere,
\begin{equation}
\label{rest-mass}
M_A  =
 \frac{(D-2)\omega_{D-2}}{16\pi G_D}\left(\frac{A}{\omega_{D-2}}\right)^{(D-3)/(D-2)}
  \ .
\end{equation}

\section{Binding energy: general results
\label{binding energy}}

\def\M{{\cal V}}
\def\n{N}
\def\Z{{\mathbb Z}_\n}
\def\N{{\cal V}/\mathbb{Z}_\n}
\def\W{{\cal W}}

Even if a brane has zero tension, it can still profoundly affect physics in the ambient
spacetime, if there is an orbifold symmetry at the brane.  Imposition of an orbifold
symmetry decimates the degrees of freedom, and thus places great restrictions
on the classical and quantum dynamics of spacetime, and of fields on spacetime.
The case of interest in this paper
is the $\mathbb{Z}_2$ orbifold symmetry of a  RS2 braneworld  
  in spacetime dimension $D=5$.  However, the issue is quite general, and also applies
to   $D>5$, as long as orbifold symmetries  hold.
In this section, we   give strong arguments that a small black hole has a large gravitational binding
energy to such a brane. 

If an asymptotically RS black hole   is sufficiently small 
  compared to the AdS curvature length   $\ell$, the values of   brane tension and   bulk cosmological constant can be neglected, since they are 
    proportional to $1/\ell$ and $1/\ell^2$, respectively (see   section \ref{RS-review} below).
By neglecting the   tension and     cosmological constant, we  will derive below a simple result for the 
 binding energy $E_B$ of a black hole  to a tensionless
brane  with   $\Z$ orbifold symmetry in
  an effectively asymptotically flat spacetime. Our derivation will assume  that  mass is  normalized with respect to an observer on the brane.
Our  result (\ref{EB-general})
  will   give accurate results for the asymptotically RS2 geometry in this paper, and it 
  can also be applied to an asymptotically RS1 geometry; we therefore regard  
  (\ref{EB-general})
  as a general result, valid for the specific case of sufficiently small black holes in the vicinity of the brane.
In section \ref{bounds}, we will obtain  upper bounds for the  mass $M$ and black hole area $A$, compared to the AdS length scale, for which the
  small black hole approximation is valid.

Let   $\M$ (the bulk spacetime) be a  Riemannian manifold with     a
  submanifold $\W$ (the brane).
Let $\Z$ be a discrete symmetry of $\M$ that leaves $\W$ fixed.  All of physics is to
be invariant under $\Z$.  Thus,
all classical solutions are symmetric under $\Z$,
and
all quantum states are invariant under  $\Z$.
We can construct classical solutions of the field equations in two ways:
(i) on a manifold $\M$  containing $N$ identical copies of the solution (before
imposition of the orbifold symmetry); or (ii) on a manifold-with-boundary or
manifold-with-conical-singularity $\N$ (after imposition of the orbifold symmetry).

Global charges in    $\M$ must be divided by $N$, when
measured in the orbifolded spacetime $\N$.  This is clear in the manifold-with-boundary
view.  In the orbifold view, it follows because our representation of physics
is $\n$-fold redundant.  For instance, an electromagnetic charge $Q$   in $\M$,
away from $\W$, must have $N$ copies in all; but Gauss's law should give
$Q$, not $NQ$, for the total charge.  If the charge   lies 
on $\W$, its $\n$ copies   coincide; but we must regard the total
charge as $Q$, not $\n Q$.  Thus, in the orbifold view, the rule
is to calculate charges by flux laws on large surfaces in $\M$, but then divide
by $\n$;  this gives the same answer as from a large $\Z$ invariant surface
in $\N$.

As measured in $\M$, a small nonrotating, uncharged, black hole has
  mass $M_0$ and   area $A_0$
  related by (\ref{rest-mass}),  
\begin{equation}
A_0 = c_D M_0^{(D-2)/(D-3)} \ ,
\end{equation}
with $c_D$ a   constant. To
construct a braneworld black hole centered on the brane, we
apply a $\Z$ symmetry, under which     the black hole  is invariant, as viewed  in $\M$.  
As measured in $\N$,
 the black hole has mass    $M_1=M_0/\n$ and
 area  $A_1=A_0/\n$.

To construct a black hole far from the brane,
it must appear in $N$ copies, so as to
be invariant under $\Z$.  We cannot construct a static solution for this situation, since
the black hole will be attracted by its image black holes under $\Z$.
However, we can construct initially static initial data  by the well known   method  of images  \cite{misner, lindquist}.
If the black
hole is sufficiently far     from the brane, the solution in a neighborhood of
the black hole will be close to the solution for a single static black hole,
so  the black hole  has mass $M_2$  and area $A_2$ 
 related by (\ref{rest-mass}),
\begin{equation}
A_2 = c_D M_2^{(D-2)/(D-3)} \ .
\end{equation}
A black hole on the brane has mass $M_1=M_0/\n$ and  
\begin{equation}
\label{area on brane}
A_1=\frac{A_0}{\n} = c_D M_1^{(D-2)/(D-3)}{\n}^{1/(D-3)} \ .
\end{equation}
 If the black hole
could leave the brane in a reversible process, the area would remain constant,
$A_2=A_1$.  Thus, for a reversible process, 
\begin{equation}
\label{M2-off-brane}
M_2 = M_1 {\n}^{1/(D-2)} \ . 
\end{equation}
If
the  process were irreversible, then $A_2>A_1$, hence $M_2 > M_1 {\n}^{1/(D-2)}$.
In either case,   $M_2 > M_1$, so energy would be required to drive the process.
The minimum binding energy $E_B$ of the black hole to the brane is thus
\begin{equation}
\label{EB-general}
	E_B = M_2-M_1 = \left[{\n}^{1/(D-2)}-1\right] M_1 \ .
\end{equation}
Note that $E_B$ is  of order  $M_1$, and hence substantial.
One can interpret this  result  in terms of an effective binding force: 
 if a black hole is off the brane, it experiences
attractive image forces from its copies in the orbifold view.
The  results (\ref{M2-off-brane}) and (\ref{EB-general})   will be borne out  
explicitly
 in the initial data that we   will construct in section \ref{ch:Misner}.

\section{The Randall-Sundrum models
\label{RS-review}}

The  Randall-Sundrum spacetimes \cite{RS1,RS2} are portions of an
 AdS spacetime, with metric
\begin{equation}
ds^2_{\rm RS} = \label{g_RS} \psi_0^2 \left(- dt^2 +
d\rho^2 + \rho^2d\omega_{D-3}^2 + dz^2 \right) \ .
\end{equation}
Here $d\omega_{D-3}^2$ denotes the unit $(D-3)$-sphere.
The    warp factor is   $\psi_0=\ell/(\ell+z)$,  where $z$  is the extra dimension
and $\ell$ is the AdS curvature length, related to the bulk cosmological constant $\Lambda$    given below.
The RS1 model \cite{RS1} contains two  branes,   the    surfaces      $z=0$ and $z=z_c$.
 The  bulk
cosmological constant $\Lambda$ and brane tensions  $\lambda_i$  are
  \begin{equation}
  \label{RS-tensions}
\Lambda    =   -\frac{(D-1)(D-2)}{2\ell^2} \ , \quad \lambda_1 =
-\lambda_2    = \frac{2(D-2)}{8\pi G_D\ell}
 \ .
\end{equation}
The dimension $z$ is   compactified on an orbifold 
($S^1/\mathbb{Z}_2$) and
the branes have    orbifold mirror
 symmetry:
in the covering space,  symmetric points across a brane are identified.  
There is a  discontinuity   in the extrinsic curvature $k_{ab}$ across each brane
 given by the Israel condition  \cite{israel}. Using
 orbifold symmetry, the Israel condition
 requires  the   extrinsic curvature $k_{ab}$ at each brane to satisfy
\begin{equation}
\label{RS-K}
2k_{ab} = \left(\frac{8\pi G_D\lambda}{D-2}\right)
\sigma_{ab} \ .
\end{equation}

The RS2 spacetime is   obtained   from RS1
 by   removing negative-tension brane (now a regulator)
 to infinite distance ($z_c \rightarrow \infty$) 
and the orbifold region has $z \ge 0$.

\section{Formulation of the initial data
\label{chap:setup-initial-value}}

\begin{figure}[h]
\includegraphics{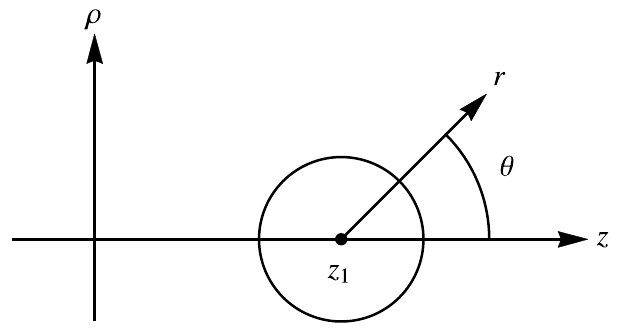}
\caption{\label{basic-coords}
The brane and black hole throat.  The brane (at left) is the plane $z=0$.  The  black hole throat  (the circle) has radius $r=a$ in    coordinates $(r,\theta)$ centered at $z=z_1$.}
\end{figure}

Here we formulate the   
initial data    for small initially static 
 black holes in RS2. As described in  
  \cite{paper-1-first-law},
  this amounts to solving the 
 constraint   
${\cal R}  = 2\Lambda$, where
 ${\cal R}$ is  the Ricci scalar  of the spatial metric $h_{ab}$.
  The spatial  geometry can be visualized as a conventional
 embedding diagram  \cite{misner, lindquist}  with   two
  asymptotically RS2 regions (instead of asymptotically flat regions),  connected
  by  a  bridge.  
  The bridge's   minimal surface  is the
apparent horizon,
which  is the best 
  approximation to   the event horizon  
within the spatial geometry.
If the black hole is sufficiently far from the brane, we refer to the  apparent horizon as the throat.
If the black hole is sufficiently close to the brane, the apparent horizon is the outermost
extremal surface surrounding the throat.  
Figure \ref{basic-coords} illustrates the  setup in  convenient coordinates.

We use a conformally flat   metric,
\begin{equation}
\label{conformal-ansatz}
ds^2 = \psi^{4/(D-3)} d{\bf x}^2 \ ,
\quad
{\bf x} = (\vec\rho,z) \ .
\end{equation}
Here  ${\bf x} = (\vec\rho,z)$ are
Cartesian coordinates, with 
  Laplacian $\nabla^2$.
The constraint is then
\begin{equation}
\label{ham-dim-d}
\nabla^2\psi
 =
 \frac{(D-1)(D-3)}{4\ell^2} \, \psi^{(D+1)/(D-3)} \ .
\end{equation}
This is
invariant under   inversion $J$ through a sphere
of radius  $a$ and center  ${\bf C}$. $J$ acts
 on
 ${\bf x}$ and   functions  $f$ as
\begin{subequations}
\begin{eqnarray}
\label{Jmap-1}
J{\bf x} &=&    {\bf C}  +
 \frac{a^2}{|{\bf x}-{\bf C}|^2} ({\bf x}-{\bf C}) \ ,
\\*
\label{Jmap-2}
 J[f]({\bf x})  &=&
 \frac{a^{D-3}}{|{\bf x}-{\bf C}|^{D-3}}  \,  f\left(J{\bf x}\right) \ .
\end{eqnarray}
\end{subequations}
 Note that $J^2$ is the identity, and
\begin{subequations}
\begin{eqnarray}
\label{J-identity-1}
 d(J{\bf x})^2 &=& \frac{a^{4}}{|{\bf x}-{\bf C}|^{4}} \, d{\bf x}^2 \ ,
 \\*
 \label{J-identity-2}
\nabla^2 J[f]  &=&
\frac{a^4}{|{\bf x}-{\bf C}|^4} \,J \left[\nabla^2  f \right]
 \ .
\end{eqnarray}
\end{subequations}
From (\ref{J-identity-2}),
 if $\psi$  is a solution of   (\ref{ham-dim-d}),
then so is $J[\psi]$.
If the inversion   is an isometry of the metric 
(\ref{conformal-ansatz}), then  
\begin{equation}
ds^2 =   [\psi(J {\bf x})]^{4/(D-3)} d({J\bf x})^2 \ .
\end{equation}
By (\ref{J-identity-1}), this requires
$J[\psi]  =  \psi$, which we will impose   across the   throat.
In the covering space, we    also impose   orbifold  reflection   isometry
about the brane.
Since the conformal  isometries of
flat space  are limited to
  reflections    and inversions,
we take the  brane as the coordinate plane $z=0$   and the throat as a coordinate 
sphere
of radius $a$ and center ${\bf C} = (\vec 0,z_1)$
in  Cartesian coordinates $(\vec\rho,z)$.
For cylindrical coordinates ($\rho,z,\varphi_i$)
and    spherical coordinates
 ($r,\theta,\varphi_i$) centered at ${\bf C}$,
\begin{equation}
|\vec\rho| = \rho = r\sin\theta \ , \quad z = z_1 + r\cos\theta \ .
\end{equation}
In  the spherical coordinates
 ($r,\theta,\varphi_i$),  the inversion isometry condition $\psi=J[\psi]$ is
\begin{equation}
\label{isometry-sph}
\psi(r,\theta,\varphi_i)  =
\left(\frac{a}{r}\right)^{D-3}  \psi(r^\prime,\theta,\varphi_i)
\ , \quad r^\prime =\frac{a^2}{r}  \ .
\end{equation}
Our two isometries are then
\begin{equation}
\label{isometry}
 J[\psi]  =  \psi \ , \quad \psi(\vec\rho,-z) = \psi(\vec\rho,z)  \ .
\end{equation}
The boundary conditions are, with   $q=(D-1)/(D-3)$,
\begin{subequations}
\label{bc-all}
\begin{eqnarray}
\label{bc-brane}
\left[
2\ell\partial_z\psi   +   (D-3)   \psi^{q}
\right] \Big|_{z=0}
&=& 0
\\*
\label{bc-throat}
\left[
 2r \partial_r\psi  +  (D-3)\psi
 \right] \Big|_{r=a}
 &=& 0
\\*
 \psi  &  \!\!\! \underset{|{\bf x}|\rightarrow\infty}{\longrightarrow}  \!\!\!  &  \psi_0 \ .
\end{eqnarray}
\end{subequations}
The result  (\ref{bc-brane}) follows from  the Israel condition (\ref{RS-K}), and (\ref{bc-throat}) follows from  differentiating
 (\ref{isometry-sph}).

Thus,
to construct an initially static  geometry  for 
a  small black hole  ($a \ll \ell$), we must solve  the
boundary value problem consisting of (\ref{ham-dim-d}), 
(\ref{isometry}),
and    (\ref{bc-all}).  We will solve this outside of the throat. Inversion   could then be used to extend the solution inside the throat, if desired.

Although  (\ref{ham-dim-d}) and
 (\ref{bc-brane}) are nonlinear,  
  this system should be
 solvable, since
$\psi$     should   interpolate
between two known solutions:
far from the apparent horizon,    $\psi$  approaches the  RS solution (\ref{g_RS}),
while near the apparent horizon,
 the AdS curvature has little effect    and
$\psi$ approaches the $D$-dimensional Schwarzschild solution.

We   will solve the above
boundary value problem  
 in the next two sections, in  two different regimes:
 in section \ref{ch:Misner}, the black hole 
is  on or near  the brane, and in
  section \ref{ch:R-S},
the black hole is
farther from the brane.

\section{Black holes near the brane
\label{ch:Misner}}

Here we solve the   boundary value problem of
section \ref{chap:setup-initial-value},
for  small   black holes ($a \ll \ell$)
 on or  near the brane.
We     obtain the solution  using a linear approximation and generalizing 
Misner's method of images \cite{misner, lindquist}
to higher dimensions.
We then compute the relevant physical
quantities.  Lastly, we apply our variational principle (\ref{vp}).  From this, we 
find
    a static black hole
on the brane, and we determine its stability and binding energy.

\subsection{Solution from method of images}

\begin{figure}[h]
\includegraphics{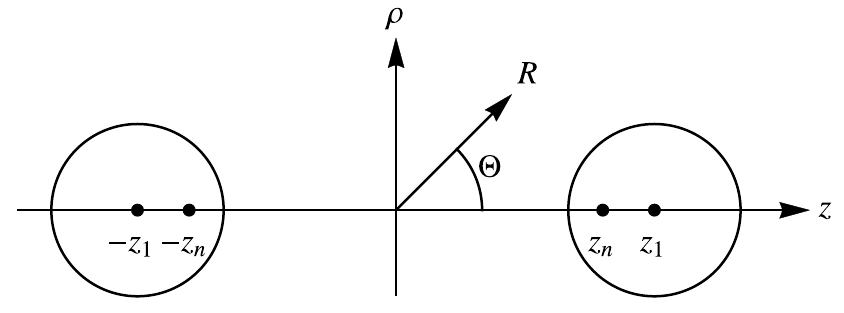}
\caption{\label{Misner-coords}
The method of images.  The black hole throat (right) and its orbifold mirror copy (left) are equidistant from the brane ($z=0$). 
The throat  coordinate radius is $a=c \, {\rm csch} \,\mu_0$. Image points at $\pm z_n$ and coordinates  ($R$, $\Theta$) are also shown.}
\end{figure}

 The boundary value problem consists of
 (\ref{ham-dim-d}), 
(\ref{isometry}),
and    (\ref{bc-all}).
For a small  black hole   sufficiently  close  to the brane ($|{\bf x}| \ll \ell$), 
   the brane tension and cosmological constant can 
be neglected, so 
we approximate $\ell \rightarrow \infty$.
This reduces (\ref{ham-dim-d})
and  (\ref{bc-all})
to an effective linear problem in an asymptotically flat space,
\begin{equation}
\label{misner-bvp}
\nabla^2\psi = 0
\ , \quad
 \partial_z \psi \Big|_{z=0} = 0
\ , \quad
 \psi  \underset{|{\bf x}|\rightarrow\infty}{\longrightarrow}  1
\ .
\end{equation}
We   implement the isometries  (\ref{isometry}) by
 using two symmetric throats ($j=1,2$) on either side of the brane ($z=0$)
and  requiring inversion isometry across
each throat, 
\begin{equation}
\label{isometry-2}
J_j[\psi]=\psi     \ , \quad
{\bf C}_1=-{\bf C}_2 =  (\vec 0, z_1)
  \ .
\end{equation}
To  construct
the metric outside the throats,
we now  generalize
Misner's method of images \cite{misner, lindquist} to higher dimensions $D$.  The procedure is 
illustrated in 
Fig.\ \ref{Misner-coords}.

From (\ref{Jmap-2}), we first note that
$J_j[1]$     is
a pole at ${\bf C}_j$ of strength $a^{D-3}$,
and  the action of $J_j$ on a pole (at $\bf y$ of strength $q$)  is
a pole at $J_j{\bf y}$ of strength
\begin{equation}
\label{imagecharge}
 q^\prime  =  q \, \frac{ a^{D-3}}{|{\bf y}-{\bf C}_j|^{D-3}} \ .
\end{equation}
We   solve (\ref{misner-bvp})\textendash(\ref{isometry-2}) by  an infinite series
$
\psi  =  S[1]
$,
where
\begin{equation}
\label{series2} S = 1  + \sum_{n=1}^{\infty} \left[ (J_1 J_2
J_1 \cdots J_{i_n})  +  (J_2 J_1 J_2 \cdots J_{i^\prime_n})
\right] \ .
\end{equation}
We can easily verify  that $J_j[S]=S$, which guarantees that  $\psi$
satisfies the inversion isometry in  (\ref{isometry-2}).
Since each term
in parentheses yields  a pole,
\begin{equation}
\label{psi-misner}
\psi = 1  + \sum_{n=1}^{\infty} q_n
\left(
\frac{1}{|{\bf x}-{\bf x}_n|^{D-3}}
 +
\frac{1}{|{\bf x}+{\bf x}_n|^{D-3}}
\right) \ ,
\end{equation}
where the poles  at $\pm {\bf x}_n=(\vec 0,\pm z_n)$ lie inside the throats, and
  ${\bf x}_1={\bf C}_1$.  By reflection symmetry,
poles at $\pm {\bf x}_n$  have
equal coefficients   and
 $\partial_z \psi =0$ at the brane.
 Since ${\bf x}_n = J_1(-{\bf x}_{n-1})$
    we have
    from (\ref{Jmap-1}) and (\ref{imagecharge}), respectively,
\begin{equation}
 z_n = z_1  -   \frac{a^2}{z_1 + z_{n-1}}
\ , \quad
\frac{q_n}{q_{n-1}}=\left(\frac{a}{z_1+z_{n-1}}\right)^{D-3} \ ,
\end{equation}
with  solutions   (easily proved  by induction)
\begin{equation}
\label{zn and qn}
z_n = c \coth n\mu_0
\ , \quad
  q_n = (c \, {\rm csch} \, n\mu_0)^{D-3}
\ .
\end{equation}
Here 
  $a=c \, {\rm csch} \,\mu_0$,
where
  $\mu_0$ is a dimensionless measure
  of the throat-brane separation,
  and $c$ is a   scale parameter.
  As in \cite{lindquist},
  bispherical coordinates ($\mu, \eta$)
are   defined by
\begin{equation}
\label{bispherical}
\tanh\mu  =  \frac{2cz}{\rho^2+z^2+c^2}
\ , \quad
\tan\eta  =  \frac{2c\rho}{\rho^2+z^2-c^2} \ .
\end{equation}
In these coordinates,
the throats  are the surfaces
  $\mu=\pm\mu_0$,
and the brane is the surface $\mu=0$.
Also, the line joining the foci 
$(\vec\rho,z)=(\vec 0,\pm c)$ is $\eta=\pi$.
In bispherical coordinates, the metric    is
\begin{equation}
ds^2 = \Phi^{4/(D-3)}  c^2 \left( d\mu^2  +
d\eta^2 + \sin^2\eta \,d\omega_{D-3}^2  \right) \ ,
\end{equation}
where, with   $\nu = (D-3)/2$,
\begin{equation}
\label{misner-Phi-1}
\Phi = \sum_{n=-\infty}^{\infty}
[ \cosh(\mu+ 2n\mu_0)-\cos\eta  ]^{-\nu} \ ,
\end{equation}
This can be expanded in  Gegenbauer polynomials $C^\nu_j$,
\begin{equation}
\label{misner-Phi-2}
\Phi= \sum_{n=0}^{\infty}
\sum_{j=0}^\infty \frac{2^{(D-1)/2}}{e^{\mu(2n+1)(j+\nu)}}  \,C^\nu_j(\cos\eta) \ .
\end{equation}
Another useful coordinate system is provided by
 spherical coordinates centered at ${\bf x}=0$,
\begin{equation}
\label{misner-R-coords}
ds^2
= \psi^{4/(D-3)} \left[ dR^2  +  R^2 \left(d\Theta^2 +
\sin^2\Theta \,d\omega_{D-3}^2 \right) \right] \ ,
\end{equation}
where for $R>z_n$  we have the multipole expansion
\begin{equation}
\label{misner-psi-multipole}
\psi =1  +
 \sum_{n=1}^\infty \frac{2q_n}{R^{D-3}}
 \left[
1
   +    \sum_{k=1}^\infty
\left(\frac{z_n}{R}\right)^{2k}
C^\nu_{2k}(\cos\Theta)
\right] \ .
\end{equation}
Also, by (\ref{bispherical}),   the throat surface
$R_{t}(\Theta)$ is the solution to
\begin{equation}
\label{misner-R-coords-throat}
R_{t}^2  - (2 c \coth \mu_0\cos\Theta) R_{t}  +   c^2  =0 \ .
\end{equation}

\subsection{Physical properties and binding energy}

We now compute the   physical
quantities   needed
to apply our variational principle  (\ref{vp}) and calculate the 
 binding energy.
The   throat-brane separation    is
\begin{equation}
L = c \int_{0}^{\mu_0} d\mu\, [ \Phi(\mu,\pi)]^{2/(D-3)}
\ .
\end{equation}
For   $D=5$, we find from
(\ref{misner-Phi-1})
\begin{equation}
 L =c  \sum_{n= -\infty}^{\infty}
\left[ \tanh (n+\mbox{$\frac{1}{2}$})\mu_0   -\tanh (n-\mbox{$\frac{1}{2}$})\mu_0  \right] \ .
\end{equation}
Evaluating the sum using $\lim_{N \rightarrow \infty} \sum_{-N}^N$  yields  simply
\begin{equation}
 \label{length-misner}
 L = 
 c \ .
\end{equation}
From the monopole term in
  (\ref{misner-psi-multipole}), the mass is
\begin{equation}
\label{mass-misner}
 M =
   \frac{(D-2)\omega_{D-2}}{4\pi G_D}
   \sum_{n=1}^\infty q_n \ ,
\end{equation}
with $\omega_{D-2}$   the area  of the unit $(D-2)$-sphere.
One finds     $L^{D-3}/(G_D M)$
 is an increasing function of
  $\mu_0$, so  $\mu_0$ is a dimensionless measure
  of the throat-brane separation, as stated earlier.
For sufficiently large  $L$,
  the apparent horizon is the throat, whose area $A_t$ is
\begin{equation}
\label{Athroat}
A_t =
c^{D-2}  \omega_{D-3}\int_0^\pi d\eta\,
 (\sin\eta)^{D-3} \Phi(\mu_0,\eta)^p \ ,
\end{equation}
where   $p=2(D-2)/(D-3)$.
For sufficiently small $L$,
the apparent horizon  $R(\Theta)$ is
the outermost extremal surface,
   surrounding the throat (\ref{misner-R-coords-throat}) and intersecting the brane.
The   area  functional   is
\begin{equation}
\label{area-outer}
 A  = \omega_{D-3}\int_0^\pi d\Theta\
(\sin\Theta)^{D-3}  R^{D-3}
\sqrt{R^2+ \dot{R}^2}\, \psi^p \ ,
\end{equation}
where $\dot R \equiv dR/d\Theta$.  Extremizing  this area $A$ gives
\begin{eqnarray}
\nonumber
\frac{R+\ddot{R}}{1+\dot{R}^2/R^2} &=&
-  \dot{R}
\left[
(D-3)\cot\Theta + p\, \frac{\partial_\Theta\psi}{\psi}
\right]
\\
\label{AH-PDE}
& &
 + R \left[ (D-1)  +  p\, \frac{R\partial_R \psi}{\psi}\right] \ .
\end{eqnarray}
At $\Theta=0$, this becomes (by L'H\^{o}pital's rule)
\begin{equation}
2\ddot{R}  =
 (D-2) R    +       p R^2 \frac{\partial_R\psi}{\psi}
\ .
\end{equation}
To find extremal surfaces surrounding the throat, we
 numerically integrate (\ref{AH-PDE}),
with the initial conditions  $R(0)>R_t(0)$
and
$\dot R =0$.
If $\dot R=0$ at $\Theta=\pi/2$,
then $R(\Theta)$ is one of
  two extremal surfaces, as also occurs in Brill-Lindquist initial data
 \cite{bishop}.  The   outermost extremal surface is
 the  apparent horizon,  
 with area $A_o$ given by (\ref{area-outer}).
We  numerically find that   such extremal surfaces surround the throat
   for $\mu_0 \le \bar\mu_0$,
where $\bar\mu_0 \simeq 1.36$,  $0.75$,  $0.51$
for $D=4$, $5$, $6$, respectively.

To apply our variational principle  (\ref{vp}),
  we   extremize the mass $M$ while holding the apparent horizon area $A$ constant.
Since $c$ is a scale parameter,  $A   =  c^{D-2}  \hat{a}$,
 where $\hat{a}$ is  dimensionless. 
 For a constant value $A$, we thus set
\begin{equation}
\label{misner-c-function}
c(\mu_0)  =  \left[\frac{A}{\hat{a}(\mu_0)}\right]^{1/(D-2)}
\ , \quad
\hat{a}   =  \left\{
\begin{array}{ll}
 \hat a_o &    \mbox{if $\mu_0 < \bar\mu_0$}
\\
 \hat a_t &   \mbox{if $\mu_0 > \bar\mu_0$}
\end{array}
\right.
\end{equation}
Here  
 $\hat a_{o}$ is found numerically as described above,  and  
 \begin{equation}
\label{athroat}
\hat a_{t}(\mu_0) =
   \omega_{D-3}\int_0^\pi d\eta\,
 (\sin\eta)^{D-3} \Phi(\mu_0,\eta)^p \ .
\end{equation}
The mass $M$
at fixed area is  then given by (\ref{mass-misner}), using $c$ in
 (\ref{misner-c-function}) to
evaluate   the coefficients  $q_n$ in (\ref{zn and qn}).
For $D=5$, we have $L=c(\mu_0)$ by (\ref{length-misner}).  For
 large throat-brane
separation ($\mu_0 \gg 1$),    
 \begin{equation}
 \frac{G_5 M}{6\pi c^2} \simeq    e^{-2\mu_0} + 3 e^{-4\mu_0}
  \ , \quad
 \frac{\hat a_t}{128\pi^2} \simeq    e^{-3\mu_0} + 6 e^{-5\mu_0}  \ .
\end{equation}
 Combining these   gives, for large  throat-brane
separation,
\begin{equation}
\label{interaction-energy-misner}
M \simeq   M_A   -
 \frac{G_5 M_A^2}{6\pi L^2} \ .
\end{equation}
This has the physically  expected form: the first term is the
 rest mass (\ref{rest-mass}), and the second term is the  
interaction energy  of the black hole with its  orbifold image.

\begin{figure}[h]
\includegraphics{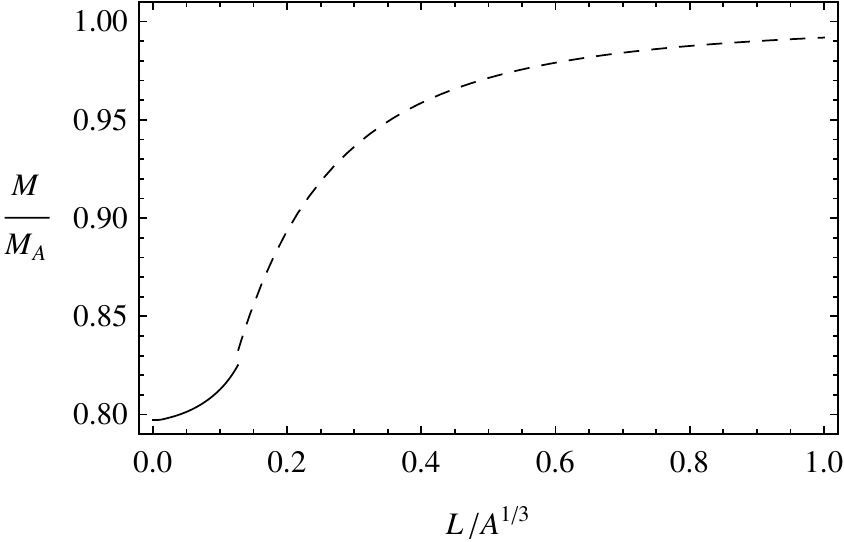}
\caption{\label{MassMisnerFigure}
Mass $M$ at fixed  area $A$,
for $D=5$.
Solid line:
the black hole is   on the brane ($A$ is the area of the outermost extremal surface).
Dashed line: the
black hole is off the brane ($A$ is the throat area).}
\end{figure}

The    mass   $M$ as a function of $L$ at fixed apparent horizon area
 is plotted
in   Fig.\ \ref{MassMisnerFigure}  for $D=5$.
By our variational principle
 (\ref{vp}), the extremum
 at $L\rightarrow 0$
  is a static black hole.
This black hole  on the brane is well known  numerically
\cite{Kudoh-smallBH-1,Kudoh-smallBH-2}. Our results   
show that this black hole
is a local mass
minimum,
which indicates it is
 stable against translations
 transverse to the brane.
 
The  black hole's stability is also  indicated by 
  its large   binding energy.
At $L\rightarrow 0$,   the     value    $M\rightarrow M_1$
in Fig.\ \ref{MassMisnerFigure}
is the value 
 (\ref{M2-off-brane})
 we  found previously in  deriving our   binding energy result  (\ref{EB-general}).
For $\mathbb{Z}_2$ orbifold symmetry,   this value is
 $M_1 = 2^{-1/3} M_A \simeq 0.79 M_A$, where the mass far from the brane is 
  $M_2 \simeq M_A$ by (\ref{interaction-energy-misner}).
  From our  binding energy result (\ref{EB-general}),
the  minimum binding energy   is  
\begin{equation}
\label{EB-misner}
E_B =
\left( 2^{1/3}-1\right) M_1 \ .
\end{equation}
This binding energy is  of order  $M_1$, and hence substantial.
Since  brane tension and cosmological constant have been neglected here,
this result explicitly demonstrates    the significant role 
of the  orbifold symmetry in  
binding  the black hole to the brane, as discussed in section \ref{binding energy}.

\section{Black holes far from the brane
\label{ch:R-S}}

\begin{figure}[h]
\includegraphics{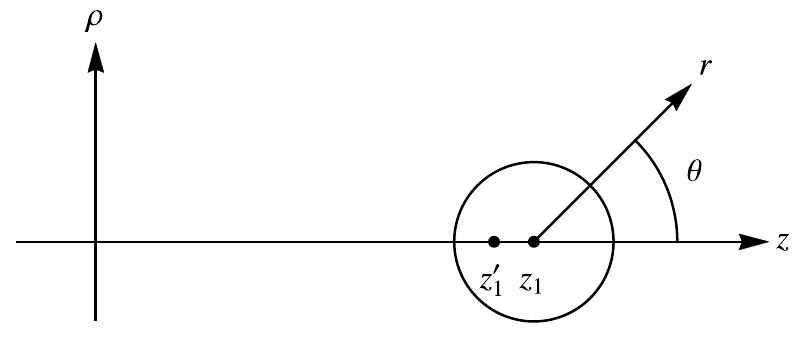}
\caption{\label{RS-coords}
A black hole farther from the brane than in section \ref{ch:Misner}.
The coordinates $(\rho,z)$ and $(r,\theta)$ are the same as in Fig.\ \ref{basic-coords}.
The source point at $z_1^\prime$   is  used in the solution  for the  field $\phi_1$.}
\end{figure}

In this section, we
solve the    boundary value problem of
section \ref{chap:setup-initial-value},
for a  small   black hole  ($a \ll \ell$, $a \ll z_1$)  
 farther from the brane than   in section \ref{ch:Misner}. 
The setup is illustrated in Fig.\
\ref{RS-coords}.
For $D=5$, we  will first   construct the asymptotically RS2 solution for $\psi$ 
using a field expansion, 
\begin{equation}
\label{phi-expansion-1}
\psi = \psi_0 + \psi_0^2  \,\phi_1 + \psi_0^2 \sum_{i \ge 2}\phi_i \ .
\end{equation}
The    field $\phi_1$ provides all of our 
  physical results.
After computing the relevant physical
quantities, we  then apply our variational principle  (\ref{vp})
 to find a static black hole
off the brane.
We also find small corrections to the binding energy (\ref{EB-misner}).
Lastly, we examine the perturbations $\phi_i$ ($i\ge 2$)
 to ensure that the   
  expansion (\ref{phi-expansion-1}) is well controlled.

\subsection{Perturbation series}

The boundary value problem for the field $\psi$
consists of
 (\ref{ham-dim-d}), 
(\ref{isometry}),
and    (\ref{bc-all}).
To solve this, we    define a related
    field $\phi$
  and two operators   (${\cal H}$, ${\cal D}$) by
\begin{eqnarray}
\nonumber
\psi &=& \psi_0 + \psi_0^2\phi
\\*
\nonumber
{\cal H} f  &=&
\psi_0^2 \left(\nabla^2  - \frac{4\psi_0}{\ell}\ \partial_z\right)f
\\*
{\cal D}f &=&
-\left[r \partial_r  + \frac{\ell+z_1-r\cos\theta}{\ell+z_1+r\cos\theta}\right]f
\ .
\end{eqnarray}
The   constraint (\ref{ham-dim-d}) is then
\begin{equation}
\label{ham-dim-phi}
{\cal H}\phi =
\frac{2}{\ell^2}\psi_0^5\left(3 \phi^2 +  \psi_0  \phi^3\right)
\end{equation}
and the boundary conditions  (\ref{bc-all}) are
\begin{equation}
\label{phi-all}
 \left(\!\partial_z \phi  + \frac{\phi^2}{\ell}   \right)\! \Big|_{z=0}\!\! =0
\ , \quad
  {\cal D}\phi  \Big|_{r=a} \!\! = \frac{1}{\psi_0(z_1)}
  \ ,
 \quad
\phi \! \underset{|{\bf x}|\rightarrow\infty}{\longrightarrow} \! 0
\ .
\end{equation}
We now write $\phi$ as a perturbation series,
\begin{equation}
\label{phi-expansion}
\phi=  \phi_1 +  \sum_{i \ge 2}\phi_i \ .
\end{equation}
All of our 
  physical results will be due to the primary field   $\phi_1$.  The fields
  $\phi_i$ with $i \ge 2$ are perturbations.
We substitute 
(\ref{phi-expansion})
into
 (\ref{ham-dim-phi})\textendash(\ref{phi-all}), 
 and
collect  terms of order $i$.  We also introduce 
parameters  $\alpha_i$ for the
 throat boundary condition, 
 obeying $\sum_{i}  \alpha_i = 1$, and we let $aA_i = \alpha_i/\psi_0(z_1)$.
This results in the following boundary value problem to solve at each order $i$,
\begin{equation}
{\cal H}\phi_i = F_i \ , \quad
\label{phi-i-bc}
 \partial_z \phi_i \Big|_{z=0} \!\! = -B_i\Big|_{z=0}
\ , \quad
 {\cal D}\phi_i \Big|_{r=a} \!\! =  aA_i
\end{equation}
and  $\phi_i \rightarrow 0$ as $|{\bf x}|\rightarrow\infty$.
The sources  $F_i$ and $B_i$
involve    only fields $\phi_j$ with $j<i$.
For the primary field $\phi_1$,
\begin{equation}
F_1  =   0
\ , \quad
B_1 = 0
\ .
\end{equation}
At second order,
\begin{equation}
\label{sources-2}
 F_2  =   \frac{6}{\ell^2} \psi_0^5 \phi_1^2
 \ , \quad
B_2 =  \frac{1}{\ell} \psi_0^2 \phi_1^2
 \ .
\end{equation}
At third order,
\begin{equation}
\label{sources-3}
F_3 =
 \frac{2}{\ell^2}\psi_0^5 \left(  \psi_0\phi_1^3 + 6\phi_1 \phi_2 \right)
\ , \quad
B_3 = \frac{2}{\ell} \psi_0^2 \phi_1\phi_2
\ ,
\end{equation}
and this process    continues to higher orders.

\subsection{Green function ${\cal G}_N$ and primary field $\phi_1$ }

For  the operator ${\cal H}$,
  the Neumann Green function ${\cal G}_N$ 
can be regarded as the field
of a  point source, defined  by
\begin{equation}
\label{GN-properties}
{\cal H} {\cal G}_N({\bf x},{\bf x}^\prime)   =    -   \delta({\bf x}- {\bf x}^\prime)
\ , \quad
\partial_z {\cal G}_N \Big|_{z=0} \!\! =  0 \ .
\end{equation}
${\cal G}_N$ vanishes as $|{\bf x}|\rightarrow\infty$, with no throat boundary condition.
This can be solved with a Fourier method: we set 
$\delta({\bf x}-{\bf x}^\prime) = \delta(\vec\rho - \vec\rho^{\,\prime}) \delta(z-z^\prime)$
and
\begin{subequations}
\label{Fourier}
\begin{eqnarray}
\delta(\vec\rho - \vec\rho\,^{\prime}) 
&=&
\frac{1}{(2\pi)^3}\int_{-\infty}^\infty  d^3 k \, e^{i \vec k \cdot (\vec\rho - \vec\rho^{\,\prime}) }
\ ,
\\
\label{GN-integral}
{\cal G}_N({\bf x},{\bf x}^\prime)   &=&
\frac{1}{(2\pi)^3}\int_{-\infty}^\infty  d^3 k \, e^{i \vec k \cdot (\vec\rho - \vec\rho^{\,\prime}) }
F
\ ,
\end{eqnarray}
\end{subequations}
where $F(\vec k, z, z^\prime)$ is to be determined.
Substituting (\ref{Fourier})
 into (\ref{GN-properties}) 
 yields a differential equation for $F$,  whose
solution involves modified Bessel functions, $I_{5/2}$ and $K_{5/2}$.
Performing the integral (\ref{GN-integral}) then
yields the result
\begin{eqnarray}
\nonumber
 4\pi^2  {\cal G}_N({\bf x},{\bf x}^\prime)
&=&
\frac{\psi_0(z)^{-2}}{|{\bf x}-{\bf x}^\prime|^2}
 +  \frac{\psi_0(z)^{-2}}{|{\bf x}-\widetilde{\bf x}^\prime|^2} 
    \\*
& &
+
   c_1
  \ln \frac{|{\bf x}-{\bf x}^\prime|}{|{\bf x}-\widetilde{\bf x}^\prime|}
 + \psi_0(z^\prime)^2{\cal J} \ .
  \label{GN_exact}
\end{eqnarray}
The first two terms are expected:
a   source   at ${\bf x}^\prime=(\vec \rho\,^\prime,z^\prime)$
and a    source  at the  orbifold image point
$\widetilde{\bf x}^\prime=(\vec \rho\,^\prime,-z^\prime)$.
The additional terms are
\begin{equation}
{\cal J} =
\frac{6\tan^{-1}[R/(z+z^\prime)]}{\ell R}
+ \frac{3 z z^\prime}{\ell^4}
+  c_2 \, {\rm Re} \left[ ie^{\xi} \Gamma(0,\xi)\right]
\label{GN_exact-2}
\end{equation}
where
\begin{equation}
R=|\vec\rho-\vec\rho\,^\prime| \ ,\quad
 \xi =  (z+z^\prime+iR)/\ell
\end{equation}
and
\begin{eqnarray}
\nonumber
c_1 &=&  \psi_0(z^\prime)^2\frac{3}{2\ell^4}\left[R^2+(\ell+z)^2+ (\ell+z^\prime)^2\right] \ ,
\\*
c_2 &=&
-\frac{2}{\ell^5 R}(\ell^2-\ell z + z^2)(\ell^2-\ell z^\prime + {z^\prime}^2)
 \ .
\end{eqnarray}
For $\bf x$ near ${\bf x}^\prime$, the first term in
(\ref{GN_exact}) dominates, giving the
$1/|{\bf x}-{\bf x}^\prime|^2$
 behavior
  of 5-dimensional gravity.
At large  $R$, the first term in (\ref{GN_exact-2})
dominates, giving the $1/R$ behavior
  of 4-dimensional gravity, which   is the  RS2 phenomenon of  localized gravity.
At large  $R$  and large $z$,
 respectively,
\begin{equation}
\label{GN_asympt}
{\cal G}_N   \simeq
\frac{3\psi_0(z^\prime)^2}{4\pi\ell R}
     \ , \quad
{\cal G}_N  \simeq
\frac{4\psi_0(z^\prime)^2}{\pi^2\ell z} \ .
\end{equation}
We may construct the   field   $\phi_1$  from ${\cal G}_N$ by
taking ${\bf x}^\prime$   inside the throat and forming a multipole expansion
\begin{equation}
\phi_1({\bf x})   =  \sum_{n=0}^\infty f_n
  \left(\partial_{z^\prime}\right)^n
{\cal G}_N({\bf x},{\bf x}^\prime)
\Big|_{{\bf x}^\prime={\bf x}_1^\prime} \ .
\end{equation}
We     approximate
this sum as  an off-center  point source,
\begin{equation}
\label{phi1_exact}
\phi_1 \simeq f_0 \, {\cal G}_N({\bf x},{\bf x}_1^\prime)
\ , \quad
{\bf x}_1^\prime=(\vec 0,z_1^\prime)
\ , \quad z_1^\prime  =  z_1-d    \ .
\end{equation}
We determine  $f_0$ and $d$   by the throat boundary condition.
 For $z_1\ll \ell$, we use
the  first   two terms  in (\ref{GN_exact}) with
\begin{equation}
\phi_1
\simeq
\frac{f_0}{4\pi^2 \psi_0^{2}}
\sum_{k=0}^1
\frac{C_k^1(\cos\theta)}{(-1)^k}
\left[
\frac{d^k}{r^{2+k}} +
\frac{r^k}{w^{2+k}}
\right] \ ,
\end{equation}
where
$\psi_0= \ell/(\ell+z_1 + r\cos\theta)$
and $w=2z_1-d \simeq 2z_1$.
 For $z_1\gg \ell$,
it suffices to  use the
first   and third terms  in (\ref{GN_exact}), with
\begin{eqnarray}
\nonumber
\phi_1
&\simeq&
\frac{f_0}{4\pi^2 \psi_0^{2}}
\sum_{k=0}^1
\frac{C_k^1(\cos\theta)}{(-1)^k}
\frac{d^k}{r^{2+k}}
\\* & &
+
\frac{3f_0}{4\pi^2\ell^2}
\left(1 +   \frac{r \cos\theta  +d}{\ell+z_1}
\right)\ln\left(\frac{r}{2z_1} \right)\ .
\end{eqnarray}
Linearizing the throat boundary condition
in $\cos\theta$
gives
\begin{equation}
\label{f0}
f_0
 \simeq
\alpha_1 4\pi^2 a^2 \psi_0(z_1)  \left(1+\varepsilon\right) \ .
\end{equation}
For $z_1 \ll \ell$, we find
\begin{equation}
\varepsilon=\left(\frac{a}{2z_1}\right)^2
\ , \quad
d \simeq
a \,\psi_0(z_1)\left(\frac{a}{2\ell} + \frac{a^3}{8z_1^3}\right) \ .
\end{equation}
For $z_1 \gg \ell$, we find
\begin{equation}
\varepsilon= 3 \left[\psi_0(z_1)\frac{a}{\ell}\right]^2 \ln \left(\frac{a}{2z_1}\right)
\ , \quad
d \simeq
\frac{a^2\psi_0(z_1)}{2\ell} \left(1 -\frac{\varepsilon}{2}\right)
\ .
\end{equation}

\subsection{Physical properties and binding energy}

We now compute the   physical
quantities   needed
to apply our variational principle  (\ref{vp}) and calculate the 
 binding energy.
 The relevant quantities   are the throat-brane separation $L$, the throat area $A$, and the mass $M$.
In this section, we are neglecting the perturbation fields $\phi_i$ ($i \ge 2$), so 
 $\alpha_1=1$ is 
 the only throat boundary condition parameter needed here.
The   throat-brane separation   is
\begin{equation}
\label{RS proper length}
L = \int_0^{z_1-a} dz \, \psi
 \simeq
\int_0^{z_1} dz \, \psi_0(z)
 =  -  \ell \ln \left[ \psi_0(z_1) \right] \ .
\end{equation}
We now evaluate the throat area $A$.
Near the throat,  
\begin{equation}
\psi  \simeq \psi_0(z_1)
  +
\frac{f_0}{4\pi^2a^2}(1+\varepsilon)
\simeq 2\psi_0(z_1)(1+\varepsilon) \ .
\end{equation}
This gives the throat
area $A = 2\pi^2 a^3 \psi^3$. It will be
  convenient to use the related 
 rest mass $M_A$, which by (\ref{rest-mass}) is given by
$G_5 M_A = (3\pi/8) a^2 \psi^2$.
Solving  this for $a$ gives 
\begin{equation}
\label{a-throat-radius}
a^2 \simeq
\frac{2G_5 M_A}{3\pi\psi_0(z_1)^2} 
\left(1- \varepsilon^\prime\right)\ ,
\end{equation}
where for $z_1 \ll \ell$ and  $z_1 \gg \ell$, respectively,
\begin{equation}
\varepsilon^\prime =
\frac{G_5 M_A}{3\pi z_1^2 \psi_0(z_1)^2}
\ ,\quad
\varepsilon^\prime =
\frac{2G_5 M_A}{\pi \ell^2}
\ln \left[\frac{G_5 M_A}{6\pi z_1^2\psi_0(z_1)^2} \right] \ .
\end{equation}
We defined the mass $M$ for an asymptotically RS geometry   in \cite{paper-1-first-law},
which for the large $\rho$ asymptotics 
(\ref{GN_asympt})   yields   
\begin{equation}
M
=
\frac{3}{8\pi G_5} \psi_0(z_1^\prime)^{2}   f_0 \ .
\end{equation}
Using (\ref{f0}) and   (\ref{a-throat-radius}), this can be rewritten as
\begin{equation}
\label{mass}
M
\simeq
\psi_0(z_1)M_A
 -
 \psi_0(z_1)M_A
 \frac{\varepsilon^\prime}{2}
  \ .
\end{equation}
The first term is the
redshifted  rest mass,  
and the second term is 
 the  
interaction energy  of the black hole with its orbifold image.
The   redshift factor $\psi_0(z_1)$ indicates the 
repelling gravitational field of the positive-tension  brane.

\begin{figure}[h]
\includegraphics{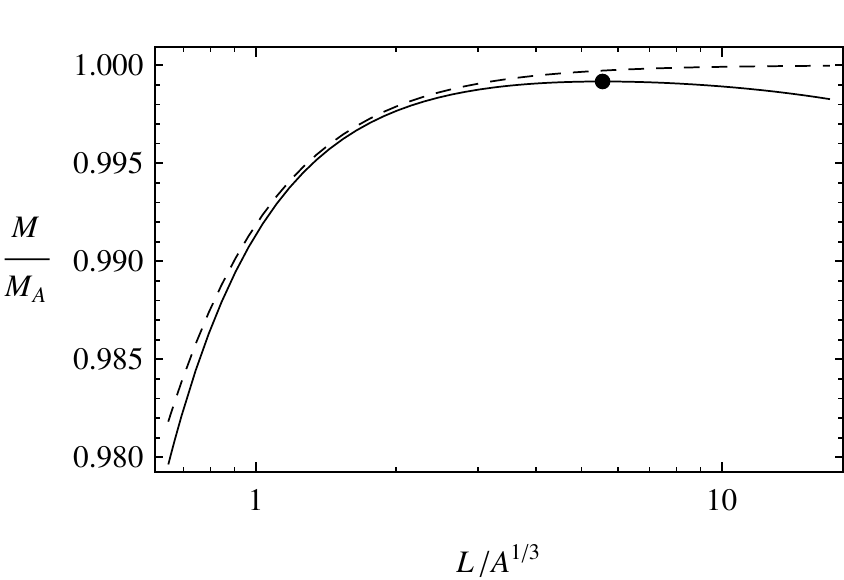}
\caption{\label{MassExtremumFigure}
Solid line: Mass $M$ at fixed  throat area $A=10^{-12}\ell^3$.
The dot denotes the 
  mass extremum.
Dashed line: the asymptotically flat approximation of Fig.\  \ref{MassMisnerFigure}.}
\end{figure}

\begin{figure}[h]
\includegraphics{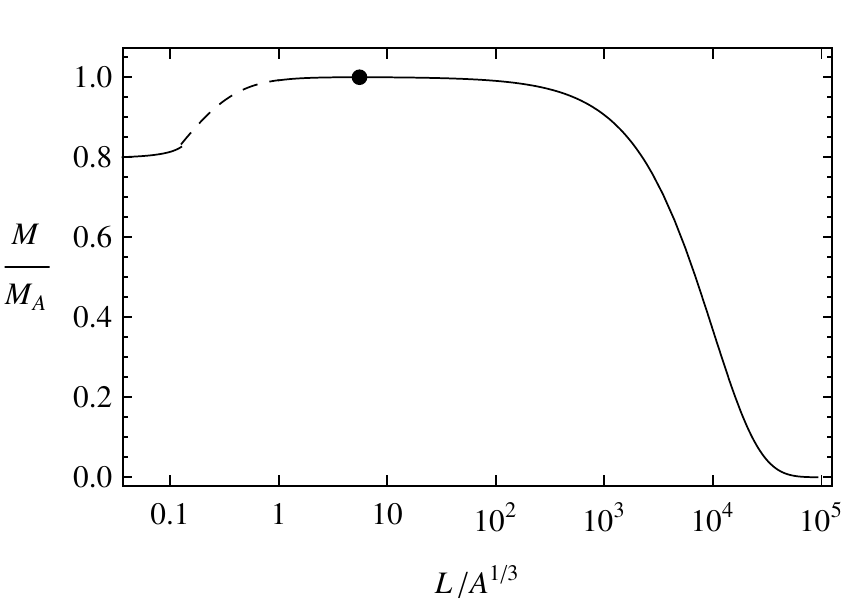}
\caption{\label{MassRedshiftFigure}
Combination of Fig.\ \ref{MassMisnerFigure}
and     Fig.\ \ref{MassExtremumFigure}.
Leftmost solid line:   the black hole is  
on   the brane.  Dashed line: the black hole is off the brane (asymptotically flat approximation). 
The dot denotes the mass extremum of Fig.\ \ref{MassExtremumFigure}.
Rightmost solid line: extension of the RS2 mass function of Fig.\ \ref{MassExtremumFigure}.
}
\end{figure}

In Fig.\ \ref{MassExtremumFigure},
we plot the  mass $M$
at fixed  area $A$.  It agrees  well
  with the
 asymptotically flat   approximation of Fig.\ \ref{MassMisnerFigure}
 in the regime
 where  the Newtonian
form (\ref{interaction-energy-misner}) is valid.
There is a  mass extremum (represented by the dot in Fig.\ \ref{MassExtremumFigure}),
 where  the black hole's repulsion
from the brane  is balanced by its  attraction to its orbifold image.
Taking   $z_1 \ll \ell$ in (\ref{mass}) gives
\begin{equation}
M \simeq  M_A   \left(1 - \frac{z_1}{\ell} \right)
\label{interaction-energy-RS-expanded}
  -  \frac{G_5 M_A^2}{6\pi z_1^2} \ .
\end{equation}
At fixed  $M_A$ and $\ell$, the
 mass extremum ($dM/dz_1=0$)   occurs at the location $z_1$ given by
\begin{equation}
(z_1)_{\rm ext}   = \left(\frac{G_5 M_A \ell}{3\pi}\right)^{1/3}
\ , \quad L_{\rm ext} \simeq (z_1)_{\rm ext} \ .
\end{equation}
By our variational principle  (\ref{vp}), this
 mass extremum    represents
a static black hole.
It is a local maximum,
  hence  this black hole is unstable against translations
 transverse to the brane.
The  binding energy   $E_B = M_{\rm ext}-M_1$   of
   a small black hole on the brane is
\begin{equation}
E_B \simeq
\left[ 2^{1/3}-1 -   \frac{3}{2}\left(\frac{2G_5 M_A}{3\pi\ell^2}\right)^{1/3} \right] M_1 \ .
\end{equation}
This is smaller than our previous estimate
  (\ref{EB-misner}),
consistent with  the   brane's repelling gravitational field.
 For large separation $L$ from the brane, 
 the    brane's  repulsion  dominates,
  and the mass at fixed area is 
  $M \simeq e^{-L/\ell} M_A$, as shown in Fig.\ \ref{MassRedshiftFigure}.

\subsection{Perturbations \label{perturbations}}

As indicated earlier, the   perturbations $\phi_i$ ($i\ge 2$) are not needed for 
our main physical results, but we consider them here   to ensure that our field 
  expansion for $\phi$ is well controlled.
The  perturbations $\phi_i$ ($i\ge 2$) 
may
be found by first constructing
a   new Green function
${\cal G}$  that  satisfies
the same relations (\ref{GN-properties}) as  the  Neumann Green function ${\cal G}_N$,
and additionally
${\cal D}{\cal G} = 0$ at the throat.
We  take
$
{\cal G}({\bf x},{\bf x}^\prime) =
{\cal G}_N({\bf x},{\bf x}^\prime) + {\cal F}_N({\bf x},{\bf x}^\prime)
$
where
\begin{equation}
\label{G-throat}
{\cal F}_N({\bf x},{\bf x}^\prime) = \sum_{n=0}^\infty  \sum_{m=1}^{(n+1)^2} f_{nm} \Gamma_{nm} \ ,
\end{equation}
and
\begin{eqnarray}
\nonumber
\Gamma_{nm}
&=&
{\cal D}_{nm}
 {\cal G}_N({\bf x},{\bf v}) \Big|_{{\bf v}={\bf v}_0} \ ,
\\*
{\cal D}_{nm} &=&   \sum_{i,j,k,l}
 S_{nm}^{ijk}
\, \partial_{v_1}^{i} \partial_{v_2}^{j}  \partial_{v_3}^{k} \partial_{v_4}^{l}
\ .
\end{eqnarray}
Here
  ${\bf v}$  lies inside the throat, so ${\cal H}{\cal F}_N=0$.
  The quantities $S_{nm}^{ijkl}$ are constants
  and
${\cal D}_{nm}$ is the order  $n$ derivative operator
that generates the spherical harmonics $Y_{nm}$,
\begin{equation}
{\cal D}_{nm}  \left(\frac{1}{4\pi^2|{\bf x}-{\bf v}|^{2}}\right)  \Big|_{{\bf v}={\bf v}_0}
=
\frac{Y_{nm}}{|{\bf x}-{\bf v}_0|^{n+2}} \ .
\end{equation}
Here
  $1 \le m \le (n+1)^2$ which motivates
  the sum in (\ref{G-throat}).  
  Letting 
   ${\bf v}_0$  denote the throat center,
it follows that
\begin{equation}
a^{n+2}\psi_0^2\Gamma_{nm}  \Big|_{r=a} =
\sum_{n^\prime, m^\prime}(\delta_{nm n^\prime m^\prime}   +   \Delta_{nm n^\prime m^\prime})
Y_{n^\prime m^\prime}
\end{equation}
with
$\Delta_{nm n^\prime m^\prime}$
 small quantities.  This equation  should
be invertible, so $\Gamma_{nm}$ are a complete set of functions
on the throat,  and
the   coefficients $f_{nm}$
can be determined by the  throat boundary condition.

The formal  solution  to  (\ref{phi-i-bc})
for the perturbation
  $\phi_i$ 
is then
\begin{eqnarray}
\nonumber
\psi_0(z)^2 \phi_i({\bf x}) &=&
-  \int   d^4\bar x \, \sqrt{\bar h}\,   {\cal G}\, \psi_0^2 F_i
+
 \int    d^3\bar x \, \sqrt{\bar\sigma}\,   {\cal G}\, \psi_0^4 B_i
\\* & &
+ \int    d^3\bar x \, \sqrt{\bar\sigma}\,   {\cal G}\, \psi_0^4 A_i \ .
\label{unregulated-phi-i-soln}
\end{eqnarray}
In these integrals,
  $\bar h$ and $\bar\sigma$ denote flat metrics, and
 only ${\cal G}(\bar{\bf x},{\bf x})$ depends on ${\bf x}$.
At large $\bar\rho$, the bulk and brane sources ($F_i$, $B_i$)  contain terms which fall off as $1/\bar\rho^2$ or $1/\bar\rho^3$.
In (\ref{unregulated-phi-i-soln}),   terms with $1/\bar\rho^2$ falloff
produce
 divergent bulk and brane integrals, and 
  terms with $1/\bar\rho^3$ falloff
 produce divergent contributions to the mass $M$.
However, these are only apparent divergences, not true divergences, since we can reformulate  
the perturbations so that no divergences occur, by setting
$\phi_i =  \widetilde\phi_i + \Phi_i$
with $\Phi_i$   a suitably chosen regulator.
The regulated perturbation $\widetilde\phi_i$ is then
\begin{eqnarray}
\nonumber
\psi_0(z)^2 \widetilde\phi_i({\bf x}) &=&
-  \int   d^4\bar x \, \sqrt{\bar h}\,   {\cal G}\, \psi_0^2 \widetilde F_i
+
 \int    d^3\bar x \, \sqrt{\bar\sigma}\,   {\cal G}\, \psi_0^4 \widetilde B_i
\\* & &
+ \int    d^3\bar x \, \sqrt{\bar\sigma}\,   {\cal G}\, \psi_0^4 \widetilde A_i \ ,
\label{regulated-phi-i-soln}
\end{eqnarray}
where      the  regulated sources     are
\begin{equation}
\widetilde F_i   =
F_i   -  {\cal H}\Phi_i
\ , \quad
\widetilde B_i  =     B_i  +  \partial_z\Phi_i
\ , \quad
 a \widetilde A_i   = a A_i  - {\cal D}\Phi_i
\end{equation}
and each regulator $\Phi_i$ must ensure the large $\rho$ falloff of
$\widetilde F_i$ and   $\widetilde B_i$
is   $1/\rho^n$ where $n \ge 4$.
Suitable  examples are
\begin{eqnarray}
\nonumber
\Phi_2 &=& \psi_0(z)\frac{(2G_5 m_1)^2}{\ell^2(\rho^2+b_1^2)} \ ,
\\*
\Phi_3 &=&
\psi_0(z)\frac{8G_5^2 m_1 m_2}{\ell^2(\rho^2+b_2^2)}
+
\psi_0(z)^2 \frac{(2G_5 m_1)^3}{\ell^3(\rho^3+b_3^3)} \ ,
\end{eqnarray}
with   $b_j$   constants.
Here   $m_i$ is the mass contribution at order $i$ to the total mass
$M=\sum_i m_i$.
One can continue choosing regulators    at higher orders, although
no regulators are needed at fourth order or higher, if one
tunes $m_i=0$
at second order and higher,
which can be achieved
by tuning the throat  boundary condition parameters
$\alpha_i$.

\section{Bounds for small black holes and related phenomenology \label{bounds}}

Below, we will
   obtain upper bounds, $M  \lesssim M_\ast$ and $A  \lesssim A_\ast$, for the   black hole mass  $M$ and area $A$, 
 for which our 
 results of sections \ref{ch:Misner} and \ref{ch:R-S} for   small  black holes in RS2  are expected to be valid.
 We consider black holes on the brane and off the brane, respectively, in sections \ref{bounds RS2} and \ref{bounds RS2 off brane}.
 In each case, we relate our bounds to phenomenology, including an application in section \ref{bounds RS1} to   RS1   and   searches for small black holes  produced at colliders. 
We   denote  the reduced Planck masses in five and four dimensions as  $M_{5}  = (8 \pi G_5)^{-1/3}$
and
$M_{Pl}  = (8 \pi G_4)^{-1/2} = 2.4 \times 10^{18}$ GeV in standard units.

\subsection{Small black holes  on the brane in RS2 \label{bounds RS2}}

In RS2, we first obtain upper bounds  ($M_\ast$, $A_\ast$)  on  black hole mass and area,  and then consider the related phenomenology.
A small  black   hole on the brane   is nearly spherical,     with bulk   area $A$ and mass $M$    related by 
 (\ref{area on brane}),
\begin{equation}
\label{area-mass for small BH}
A =  2^{1/(D-3)}  \, c_D \,   (G_D M)^{(D-2)/(D-3)} \ .
\end{equation}
Here and below, the numerical coefficients $c_n$  are  
 \begin{equation}
\label{coeff c}
   c_n = \left[\frac{1}{\omega_{n-2}} \left(\frac{16\pi}{n-2}\right)^{n-2} \right]^{1/(n-3)} \ ,
\end{equation}
with $\omega_{n-2}$ the area of the $(n-2)$-dimensional unit sphere.
The small black hole behavior (\ref{area-mass for small BH}) 
ceases to be valid
at mass and  length    scales  that  can be estimated
from the properties of larger black holes, which we  now consider.

\begin{figure}[h]
\includegraphics{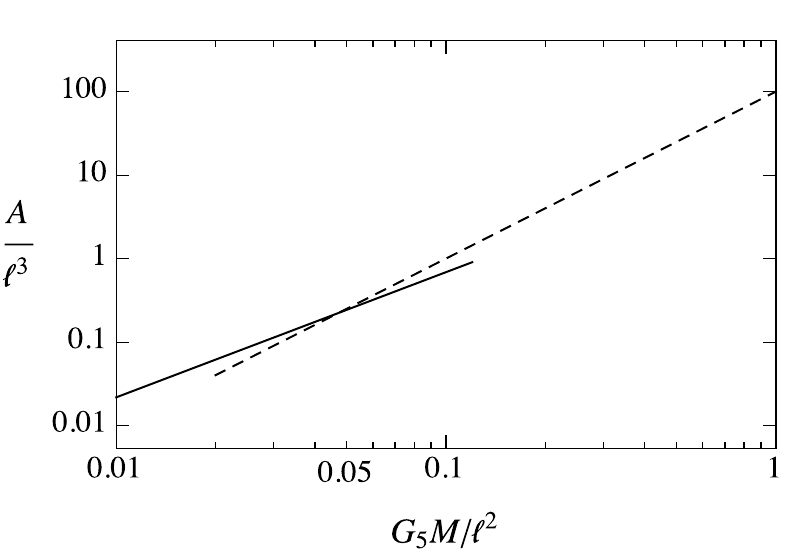}
\caption{\label{mass-ranges}
Relation between area $A$ and mass $M$ for  asymptotically RS2 black holes on the brane, in $D=5$, for small black holes (solid line, $M \lesssim M_\ast$) and large black holes (dashed line). These    intersect at   $G_5 M_\ast \simeq 0.05 \, \ell^2$. Each axis is logarithmic.}
\end{figure}

In RS2, a very large static black hole on the brane has a flattened  (pancake) shape \cite{black-string, ehm}, for which the      area $A$  in the bulk
 and   circumference ${\cal B}$  on the brane 
are  
\cite{ehm}
\begin{equation}
\label{brane area and circumf}
A = \frac{\ell \, {\cal B}}{{D-3}}  \ , \quad {\cal B}=c_{D-1} \left(G_{D-1}M\right)^{(D-3)/(D-4)}   \ .
\end{equation}
On the brane, ${\cal B}$ is a $(D-3)$-dimensional area and  
  the      Newton gravitational constant  is   
   $G_{D-1}= (D-3)G_D /{\ell}$.
Combining this with  (\ref{brane area and circumf})    
shows that  a large black hole on the brane has   bulk area $A$ and mass $M$ related by
\begin{equation}
\label{area-mass for large BH}
A = c_{D-1} (D-3)^{1/(D-4)} \left[\frac{(G_D M)^{D-3} }{\ell}\right]^{1/(D-4)} \ .
\end{equation}
We now find an upper bound on the mass  $M_\ast$ and area $A_\ast$ for which the small black hole relation  (\ref{area-mass for small BH})  is valid, by  equating the two areas  in  (\ref{area-mass for small BH}) and (\ref{area-mass for large BH}).  These     intersect at the    value    $M_\ast$ determined
 by the AdS length  $\ell$,
\begin{equation}
\label{transition mass RS2}
 G_D M_\ast =   \frac{2^{D-4}}{(D-3)^{D-3}}\left(\frac{c_D}{c_{D-1}}\right)^{(D-3)(D-4)} \ell^{D-3} \ ,
\end{equation}
with $A_\ast$ given by  (\ref{area-mass for small BH}) with $M=M_\ast$.
The result (\ref{transition mass RS2}) shows that    a  black hole  with $M \lesssim M_\ast$ is
small  compared to the   AdS curvature length scale,
    $G_D M \ll \ell^{D-3}$. 
This is illustrated     in Fig.\   \ref{mass-ranges} for
  $D=5$, for which  
$G_5 M_\ast =   (4/27)  \ell^2  \simeq  0.05 \, \ell^2$ and $A_\ast  \simeq 0.2 \, \ell^3$.

The bound  $M  \lesssim M_\ast$  is consistent with the assumption $a \ll \ell$ on the  
throat coordinate radius $a$ used  in this paper.  To see this,   
recall from
section \ref{ch:Misner}
that for small throat-brane
separation ($\mu_0\ll 1$),  the black hole apparent horizon is the 
outermost extremal surface.  As  $\mu_0 \rightarrow 0$, this surface  is a sphere, $R=R_0$,
which
can be found   analytically by taking $R \gg z_n$ in (\ref{misner-psi-multipole}),
for which 
 the monopole term in $\psi$ dominates. Solving  (\ref{AH-PDE})   yields
  \begin{equation}
  \label{R0}
R_0^{D-3} = 2\sum_{n=1}^\infty q_n  
= 2 \,  \zeta(D-3) \, a^{D-3}
\ ,
\end{equation}
where $\zeta$   is the Riemann  zeta  function.  The last equality in (\ref{R0}) follows from using
  $q_n$ in (\ref{zn and qn})
 with $c= a/{\rm csch} \,\mu_0$  and
  $({\rm csch} \, n\mu_0)/ ({\rm csch} \,\mu_0)  =1/n + O(\mu_0^2)$ as  $\mu_0 \rightarrow 0$.
The resulting   area $A$ of the  surface $R=R_0$ is   
   \begin{equation}
   \label{bound A and a}
A =  \frac{1}{2}  \left[8 \,\zeta(D-3)\right]^{(D-2)/(D-3)} \omega_{D-2}  a^{D-2}\ .
\end{equation}
The corresponding upper bound  $a \lesssim a_\ast$ on the radius $a$ is given by using the area $A_\ast$   in   (\ref{bound A and a}).
For $D=5$, 
    the value  $A_\ast  \simeq 0.2 \, \ell^3$ found above and $\zeta(2)=\pi^2/6$
gives $a_\ast \simeq 0.08 \, \ell$.  This   is   an upper bound on $a$
 for the validity of the  condition $a \ll \ell$   used  in section \ref{ch:Misner}.  

The  phenomenology of the   RS2  model is 
      currently based on  experiments that probe corrections to the  inverse square law of Newtonian gravity. 
     If our observed universe  is  indeed a brane in the RS2 model, torsion pendulum experiments  \cite{Adelberger}   yield the bound $\ell < $ 0.014 mm on the AdS curvature length $\ell$. The AdS length $\ell$ relates the
  effective Planck masses in five and four dimensions   in RS2 by  \cite{RS2}
$M_5 = (8\pi M_{Pl}^2/\ell)^{1/3}$, which  for   $\ell < $ 0.014 mm yields the lower bound $M_5 > 1.3 \times 10^6$ TeV.
This energy scale, which is relevant for the production of small black holes in particle collisions on the brane,  
is beyond the reach of current particle colliders, 
although it was suggested  in \cite{Emparan-shock-waves} that high energy signatures of RS2 
   might   be accessible by  ultra high energy cosmic rays.   Some progress was made in this direction in 
   \cite{Emparan-shock-waves},  
  by mapping out   the  different forms of 
 the cross section for producing small black holes in 
 high energy scattering experiments on the brane,   for   different ranges   of the center of mass energy and impact parameter, compared to scales set by $\ell$ and $M_{Pl}$.
Since the high energy collider phenomenology of  the RS1 model  is much more developed than that of RS2,  
 in section  \ref{bounds RS1} below, we will consider the phenomenology of RS1
and the energy range for which our  general binding energy result    (\ref{EB-general}) is applicable.

\subsection{Small black holes off the brane in RS2 \label{bounds RS2 off brane} }

The  phenomenology of RS2, relevant for black holes on the brane, has already been discussed in   section \ref{bounds RS2} above.    
Here we obtain upper bounds ($M_\ast$, $A_\ast$) on the mass and area of small black holes localized off the brane in RS2.   
We will use the results of this section when we consider
 the RS1 model in section \ref{bounds RS1} below. 

Since a small   black hole  
is nearly spherical,   the relation between its
one-dimensional circumference 
${\cal C}$ and surface area $A$ 
is well approximated by  
\begin{equation}
\label{spherical area and circumf}
{\cal C} = 2\pi    \left(\frac{A}{\omega_{D-2}}\right)^{1/(D-2)}    \ .
\end{equation}
As found in 
(\ref{RS proper length}), the black hole is located a  
proper distance  $L$ 
and
coordinate distance $z_1$
 from the brane, where
\begin{equation}
\label{L recap}
L \simeq -\ell \ln [\psi_0(z_1)] \ , \quad  
\psi_0(z_1)= \frac{\ell}{\ell+z_1}  \simeq e^{-L/\ell}
\ .
\end{equation}
Note that
$L$ essentially
  interpolates  
between the smaller  of   $z_1$ and $\ell$.  That is,
for $z_1\ll \ell$, we have  $L \simeq z_1$, while for $\ell \ll z_1$,
  we have $L \simeq \ell \ln (z_1/\ell)$, which is nearly of order $\ell$ since the logarithm is a  slowly varying function of $z_1$.
  
The condition that  the black hole is small  and   localized off the brane 
can  be stated geometrically  as    ${\cal C} \ll  L$,
hence  ${\cal C} \lesssim  L$ provides  an upper bound on ${\cal C}$ for this small black hole regime to be valid.  
Expressing this  in terms of the area $A$ using (\ref{spherical area and circumf}) then
yields the condition $A \lesssim A_\ast$, where the upper bound on the black hole area is
\begin{equation}
A_\ast = \omega_{D-2}   \left(\frac{L}{2\pi}\right)^{D-2}    \ .
\end{equation}
For $D=5$, this upper bound  is $A_\ast = L^3/(4\pi) \simeq 0.08 L^3$.  
The corresponding 
  upper bound on the mass, $M \lesssim M_\ast$, follows from evaluating
 $M \simeq e^{-L/\ell} M_A$ given in  (\ref{mass}),
  \begin{equation}
  \label{RS2 mass bound}
G_5 M_\ast  = 
    e^{-L/\ell} \left[ \frac{3\pi}{8} \left(\frac{A_\ast}{2\pi^2}\right)^{2/3} \right] \simeq 0.03 \, e^{-L/\ell} L^2   \ .
\end{equation} 
The corresponding upper bound  $a \lesssim a_\ast$ on the radius $a$ follows from evaluating
(\ref{a-throat-radius}) with $M_A \simeq M_\ast e^{L/\ell}$, 
   \begin{equation}
   a_\ast = \frac{e^{L/\ell}L}{4\pi} \simeq 0.08 \, e^{L/\ell}L \ .
 \end{equation} 
With $L$ given by (\ref{L recap}), the condition $a \lesssim a_\ast$    thus specifies a region in the space of parameters  $(a, z_1, \ell)$.
For the cases  $z_1 \ll \ell$ and  $\ell \ll z_1$, we have respectively, $a_\ast \simeq 0.08 \, z_1$ and  $a_\ast \simeq 0.08 \, z_1 \ln (z_1/\ell)$. 
These  are   upper bounds 
 on $a$
 for which the 
approach of section
 \ref{ch:R-S}   is expected to be reliable for describing an initially static 
small black hole localized off the brane.

\subsection{Small black holes  on the brane in RS1  \label{bounds RS1}}

Although  the RS2 model is the main focus of this paper, we here discuss an application  to 
   the RS1 model, which is relevant for the possible production of small black holes in LHC   experiments. 
   Our general binding energy result  (\ref{EB-general})
indicates  there is  substantial binding   of a small black hole to  a brane with orbifold symmetry.  We here obtain an upper bound on the mass for which 
this strong binding result  can be expected to apply  in the RS1 model.

In the RS1 model (see section \ref{RS-review}), our universe resides on a  negative-tension brane at a proper distance $L$ from the positive-tension brane.
An  upper bound on the mass,  $M \lesssim M_\ast$, for a black hole on the negative-tension brane to be treated as small 
 can be    obtained
  directly 
 from our previous result (\ref{RS2 mass bound}) in the RS2 model,
  \begin{equation}
  \label{RS1 mass bound}
G_5 M_\ast  =  \left(\frac{3}{64\pi} \right) e^{-L/\ell} L^2  \simeq 0.015 \, e^{-L/\ell} L^2   \ .
\end{equation} 
This value is reduced by a factor of 2 compared to (\ref{RS2 mass bound}), since the black hole resides on the brane, and 
we are measuring mass on the orbifold region between the branes.  

In the  RS1 model \cite{RS1}, the higher-dimensional Newton constant is $G_5 \simeq \ell/ (8\pi M_{Pl}^2)$. 
The RS1 model \cite{RS1} can solve the hierarchy problem if $L \simeq 12 \pi \ell$, for which (\ref{RS1 mass bound}) yields
  \begin{equation}
    \label{RS1 mass bound 2}
  M_\ast  =54 \pi^2  e^{-12\pi} M_{Pl}^2 \ell   \ .
\end{equation} 
If our observed universe    resides on the negative-tension brane in the RS1 model, 
the  current  
  phenomenological
constraints on the AdS length $\ell$ are
 \cite{LHC-RS-extra-dimensions}
   \begin{equation}
   \label{pheno bounds ell}
 10 < M_{Pl} \ell  < 100 \ ,
\end{equation} 
with
$M_{Pl} \ell$ 
 a dimensionless quantity in the standard units used here.
The lower   bound  in (\ref{pheno bounds ell}) is  due to  perturbativity requirements,  and the upper bound  is  set by high precision electroweak data.   
Combining (\ref{RS1 mass bound 2}) and (\ref{pheno bounds ell})    yields
  \begin{equation}
\label{RS1 mass bound 3}
 5.4 \times 10^2\ \mbox{TeV} < M_\ast <  5.4 \times 10^3 \ \mbox{TeV} \ .
\end{equation} 
Similar bounds to (\ref{RS1 mass bound 3}) were found in \cite{Anchordoqui} by considering ranges of coordinates, instead of  the   geometric quantities we examined above.
We thus estimate that  a black hole with mass   
$M \lesssim 5.4 \times (10^2 - 10^3)$ TeV may be reliably treated as small, with a  corresponding  large orbifold binding to the negative-tension brane as derived 
in this paper.
For a black hole produced in a collider experiment on the brane,  the   black hole mass  
is   at least of order 
  the 
higher-dimensional Planck mass.
The precise value of 
the   black hole mass 
 depends on how much energy is trapped inside the Schwarzschild radius associated with the center of mass energy.
Recent LHC collider searches  \cite{LHC-RS-black-holes}  at  center of mass energy  8 TeV   
exclude  evidence for the production of black holes with masses below 4.7\textendash 5.5 TeV, for
a  higher-dimensional Planck energy in the range 2\textendash 4 TeV in the RS1 model.  These experimental
results 
are consistent with our mass bound (\ref{RS1 mass bound 3}), which can be applied to future experimental searches for black hole production at higher center of mass energies in the TeV range or higher.

\section{Discussion \label{conclusions}}

In this paper, we
applied our
variational principle \cite{paper-1-first-law}
to initial data   for
small asymptotically RS2 black holes,   
and found two
static black holes.
We showed that the well known  static black hole
on the brane   is stable against translations transverse to the brane,  and has a large binding energy to the brane
due to  the brane's orbifold symmetry.
This is an explicit example of a simple but general  binding energy formula, given in (\ref{EB-general}), which  can be used 
in other orbifold-symmetric braneworld models.
We also found a new static black hole
off the brane, at the unique location in the bulk where
  the black hole's repulsion   from the brane  is balanced
 by its attraction to its orbifold image, with a novel instability
 to transverse translations.   
   Although the static  black hole on the brane is classically stable,
it would be  interesting to consider its quantum tunneling
through the barrier  illustrated in
  Fig.\ \ref{MassRedshiftFigure}.
It would also be interesting to study the existence and properties
of these classical solutions at larger black hole mass; numerical
methods would probably be necessary.

Our results show that a  small black hole  produced on an orbifold-symmetric brane in RS2   is stable against leaving the brane,
and we have indicated how this conclusion can be applied to models other than RS2, as long as the brane has an orbifold symmetry.
     On such a brane,  small    black holes  produced in   high energy experiments    could  be studied   directly
     (instead of leaving behind a signature of missing energy), which is an important result for future collider experiments.

\begin{acknowledgments}
  We thank  J.\  Polchinski  for useful discussions. 
     This research was supported in part by the National Science Foundation
   under Grant No. NSF PHY11-25915.
\end{acknowledgments}

\bibliography{binding2}

\end{document}